\documentclass[a4paper]{article}

\usepackage{INTERSPEECH2021}
\usepackage{multirow}
\usepackage{multicol}
\usepackage{makecell}
\usepackage[noline, boxed, linesnumbered]{algorithm2e}
\usepackage{graphicx}
\usepackage{url}

\title{Speaker Diarization using Two-pass Leave-One-Out Gaussian PLDA \\Clustering of DNN Embeddings}
\name{Kiran Karra and Alan McCree}
\address{
  Human Language Technology Center of Excellence  \\
The Johns Hopkins University, Baltimore, MD, USA}
\email{kiran.karra@jhuapl.edu, alan.mccree@jhu.edu}

\begin{document}

\maketitle
\begin{abstract}

Many modern systems for speaker diarization, such as the recently-developed VBx approach, rely on clustering of DNN speaker embeddings followed by resegmentation. Two problems with this approach are that the DNN is not directly optimized for this task, and the parameters need significant retuning for different applications. We have recently presented progress in this direction with a Leave-One-Out Gaussian PLDA (LGP) clustering algorithm and an approach to training the DNN such that embeddings directly optimize performance of this scoring method. This paper presents a new two-pass version of this system, where the second pass uses finer time resolution to significantly improve overall performance.
For the Callhome corpus, we achieve the first published error rate below 4\% without any task-dependent parameter tuning. We also show significant progress towards a robust single solution for multiple diarization tasks.
\end{abstract}

\newcommand{\zn}{{\bf z}_n}
\newcommand{\zbarML}{{\bf \bar{z}}_{ml}}
\newcommand{\mi}{{S}_i}
\newcommand{\mD}{{\bf m}_i}
\newcommand{\mz}{{\bf m}_0}
\newcommand{\cn}{{\bf c}_n}
\newcommand{\cp}{{\bf c}_{n+j}}
\newcommand{\cm}{{\bf c}_m}
\newcommand{\sn}{{\bf s}_n}
\newcommand{\on}{{\bf o}_n}
\newcommand{\op}{{\bf o}_{n+j}}
\newcommand{\om}{{\bf o}_m}
\newcommand{\SD}{{\bf \Sigma}_{i}}
\newcommand{\SML}{{\bf \Sigma}_{ml}}
\newcommand{\Sc}{{\bf \Sigma}_{wc}}
\newcommand{\Sn}{{\bf \Sigma}_{n}}
\newcommand{\Sj}{{\bf \Sigma}_{j}}
\newcommand{\Sm}{{\bf \Sigma}_{ac}}
\newcommand{\Smp}{{\bf \Sigma}^\prime_{m}}
\newcommand{\iid}{{i.i.d.}}
\newcommand{\Ulda}{{\bf U}_{lda}}

\section{Introduction}

As highlighted in the recent DIHARD challenges (2018, 2019), there are many techniques for speaker diarization~\cite{sell2018diarization, diez2018but,sun2018speaker,zajic2018zcu,vinals2018estimation, patino2018eurecom,ryant2018first,landini2020bayesian}.  Traditional approaches involve clustering of speaker embeddings (xvectors) followed by resegmentation, but we have concerns
that the DNN is not directly optimized for this task and the parameters need significant retuning for different applications. The research community has recently explored end-to-end (E2E) approaches to overcome these drawbacks~\cite{fujita2020end, huang2020speaker, medennikov2020target, kinoshita2020integrating, horiguchi2020end}. However, while E2E approaches are desirable from a philosophical perspective and can handle overlapping speech segments, they do not yet attain the best performance.

In this work, we begin with the Leave-One-Out Gaussian PLDA (LGP) approach, which clusters DNN embeddings using a Gaussian Mixture Model (GMM)~\cite{mccree2019}. In order to improve performance without task-dependent tuning, we introduce a two-pass LGP algorithm.  Here, each pass examines the audio to be diarized at different segment lengths and overlaps, to fine-tune speaker assignments for speech segments.  During each pass, the speaker assignment algorithm alternates between updating speaker models and generating speaker posteriors, while leaving out segments which were included in the model estimation to reduce bias.

The paper is organized as follows: we begin by reviewing the original LGP algorithm and DNN training process.  We then motivate and explain the two-pass algorithm in detail. Modifications needed to the LGP algorithm to accomodate varying tasks are addressed.  Finally, we discuss and compare the performance of this system against other published diarization systems on four separate datasets: 1) Callhome, 2) DIHARD2, 3) AMI Beamformed, and 4) AMI Mixed Headset.

\section{LGP Diarization}
In this section we review the LGP model as first presented in~\cite{mccree2019}.
We use the PLDA generative
model~\cite{ioffe2006probabilistic, Prince07} for clustering of DNN segment embeddings.
To tackle the problem of joint estimation of speaker models and
segment alignments, we use a leave-one-out (LOO) method to replace
the VB approach~\cite{villalba2015variational, landini2020bayesian}.
Besides giving a practical way to overcome the inherent bias of
scoring models against segments which were included in the model
estimation, this also allows us to improve performance by removing the
independence assumption of PLDA.
\subsection{Clustering}
In more detail, the LGP algorithm works as follows. Given
inputs of the length-normalized segment embeddings, initial segment
posteriors over speakers, and PLDA parameters (within-class and
across-class covariance), alternate between updating the speaker
models and generating segment speaker posteriors.

\subsubsection{Model Update}
Given the Gaussian PLDA model with known covariances $\Sc$ and $\Sm$ and a set of $N$ enrollment segments ${\bf z}_n$, we update the posterior distribution of the speaker model $\mi$ which is Gaussian~\cite{Duda01} with mean:
\begin{equation}
    \label{eq_enroll}
    \mD = \Sm\left(\Sm+\SML\right)^{-1}\zbarML
\end{equation}
and covariance:
\begin{equation}
    \label{eq_bayes}
    \SD = \Sm\left(\Sm+\SML\right)^{-1}\SML
\end{equation}
where $\zbarML=\frac{1}{N}\sum_{n=1}^{N} {\bf z}_n$ and $\SML=\frac{\Sc}{N}$ represent the maximum likelihood (ML) mean estimate and the covariance of this estimator.
To update the speaker priors (weights), we follow~\cite{kenny2008bayesian,diez2018speaker} and use a non-Bayesian maximum-likelihood approach, as it has been observed to have good properties of eliminating redundant speakers.

\subsubsection{Speaker Assignment}
The updated speaker models (means, covariances, and weights) form a
Gaussian mixture model, so speaker assignment is done by computing LOO
posteriors per class (responsibilities). LOO is implemented by leaving out the current sample, $n$, when computing GMM model parameter updates over all enrollment segments \footnote{See: \url{https://github.com/hltcoe/VBx} for more details}. Posteriors are computed using the predictive distribution, which is again Gaussian:
\begin{equation}
    \label{eq_score}
    \zn|S_i \sim {\cal N}(\mD,\Sc+\SD).
\end{equation}

\subsubsection{Dimension Reduction and Diagonalization}
To reduce computation, this work uses diagonal PLDA covariance matrices
based on the fact that two symmetric matrices can be simultaneously
diagonalized with a linear transformation~\cite{Fukunaga90}. This
process is similar to Linear Discriminant Analysis (LDA), and results in a transformed space where $\Sc = I$ and $\Sm$ is diagonal.

\subsubsection{Removing the Independence Assumption}
To better model the correlation between consecutive segments, \cite{mccree2017extended} introduced non-independent enrollment update equations. For this model, the enrollment still follows Eq.~\ref{eq_enroll} with the same ML mean, but the covariance of this ML mean estimator now decreases more slowly with increasing number of enrollment segments:
\begin{equation}
    \label{s_ml_r_oldNeff}
    \SML = \frac{\Sc}{N}\left(1+2\sum_{j=1}^{N-1}{\frac{(N-j)}{N}r^j}\right)
\end{equation}
The parameter $r$ represents the correlation between successive channel draws, and allows continuous variation between the two extremes of ``by-the-book PLDA scoring'' ($r = 0$) and ``average i-vector scoring'' ($r = 1$).

\subsubsection{Selecting Number of Speakers}
In all clustering applications, selecting the number of clusters is a
challenging problem. While AHC with PLDA comparisons does work well
for this task, it works best with a tuned, task-dependent
stopping threshold. As in \cite{diez2018speaker}, we prefer to start with a
maximum number of speakers and let the clustering algorithm
automatically select the correct number. While LGP does
produce overall likelihood estimates for any number of speakers, in
practice the ML weight updates quickly reach zero for unnecessary
speakers. We initialize the algorithm with k-means
in the diagonalized embedding space with a fixed number of speakers,
and let the algorithm eliminate speaker weights across iterations.

\subsection{DNN Training}
\label{sec_dnn_nb}
The baseline network architecture is a TDNN shown in Table~\ref{tab:xvec_nb}.
\begin{table}
  \caption{Baseline x-vector architecture. These
    experiments use a model with a layer size of 768 and an embedding dimension of 128.}
  \label{tab:xvec_nb}
  \centering
  \begin{tabular}{@{}lccc@{}}
    \toprule
    Layer & Layer Type & Context & Size \\
    \midrule
    1 & TDNN-ReLU & t-2:t+2 & L \\
    2 & Dense-ReLU & t & L \\
    3 & TDNN-ReLU & t-2, t, t+2 & L \\
    4 & Dense-ReLU & t & L \\
    5 & TDNN-ReLU & t-3, t, t+3 & L \\
    6 & Dense-ReLU & t & L \\
    7 & TDNN-ReLU & t-4, t, t+4 & L \\
    8 & Dense-ReLU & t & L \\
    9 & Dense-ReLU & t & 3*L \\
    10 & Pooling (mean+stddev) & Full-seq & 6*L \\
    11 & Dense(Embedding) & & D \\
    12 & Length-norm & & D \\
    13 & Gauss quadratic-Softmax & & Num. spks. \\
    \bottomrule
  \end{tabular}
\end{table}
This represents an extension of the x-vector
architecture~\cite{snyder2019speaker}, where the classification layers have been modified to match the diarization task. First, the traditional ReLU nonlinearity has been replaced by
length normalization, since this is well-known to improve the performance of Gaussian PLDA~\cite{Romero11}.
To implement PLDA scoring in the DNN, we add a Gaussian quadratic layer.
This layer maintains enrollment statistics
for each class using a few (1-20) past values of embeddings for that class,
and then uses the PLDA parameters $\Sc$ and $\Sm$ to perform Bayesian model enrollment using Eq.~\ref{eq_enroll} and \ref{eq_bayes}. Finally, log-likelihoods for each class are produced with the predictive distributions from Eq.~\ref{eq_score}.

The PLDA parameters are estimated in the following way. First, the within-class covariance $\Sc$ is set to identity, as we again assume the DNN can force the embeddings to match this property. We do need a discriminatively-trained scale factor to compensate for the length-normalization constraint. For the across-class covariance $\Sm$, we use a generative update of this matrix. The PLDA layer, which estimates the global PLDA parameters, maintains it's own separate copy of enrollment statistics over previous embeddings for each class, and $\Sm$ is approximated by the covariance of these ML model means.

\begin{figure}[t]
  \centering
  \includegraphics[width=\linewidth]{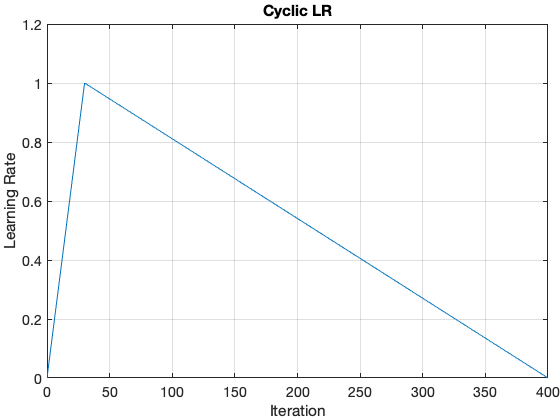}
  \caption{Learning rate schedule used for training xvector DNN.}
  \label{fig:lr}
\end{figure}

The training cost is normalized multiclass cross-entropy. To compensate for possible instability of generative parameter updates which are not exposed to gradient optimization, the learning rate uses a linear ramp-up from zero for the initial few training epochs. The learning rate schedule is shown in Fig.~\ref{fig:lr}.

\section{Algorithm Improvements}
We have a made a number of improvements to the LPG system, including replacing the TDNN with a ResNet, introducing a second pass for time refinement, and improving the duration modeling function.
\subsection{ResNet}
For wideband speech signals, we have experimented with a modified ResNet architecture shown in Table~\ref{tab:xvec_wb}~\cite{garcia2020magneto}. The Angular Margin Softmax described in ~\cite{garcia2020magneto} is replaced with a Gauss quadratic-Softmax to allow PLDA scoring within the DNN, as described in Section~\ref{sec_dnn_nb}.
\begin{table}\fontsize{6.5}{7.2}\selectfont
  \caption{Modified ResNet-34 architecture with 15.4 million parameters. Batch-norm and ReLU layers are not shown. The $1 \times 1$ convolutions are used to match the dimensions for the residual connections. The dimensions are (Channels $\times$ Frequency $\times$ Time). The input comprises 80 Melfilter bank energies from speech segments. During training we use a fixed segment length of T = 400}
  \label{tab:xvec_wb}
  \centering
  \begin{tabular}{@{}lcc@{}}
    \toprule
    Layer Name & Structure & \makecell{Output\\ $(C \times F \times T)$} \\
    \hline\hline
    Input & $-$ & $1 \times 80 \times T$ \\
    \hline
    Conv2D & $3 \times 3$, stride=1 & $128 \times 80 \times T$ \\
    \hline
    ResBlock-1 & $\begin{bmatrix} 3 \times 3, 128\\ 3 \times 3, 128 \end{bmatrix} \times 3$, stride=1 & $128 \times 80 \times T$ \\
    \hline
    ResBlock-2a & $\begin{bmatrix} 3 \times 3, 128\\ 3 \times 3, 128\\ 1 \times 1, 128 \end{bmatrix} \times 1$, stride=2 & $128 \times 40 \times T/2$ \\
    ResBlock-2b & $\begin{bmatrix} 3 \times 3, 128\\ 3 \times 3, 128 \end{bmatrix} \times 3$, stride=1 & $128 \times 40 \times T/2$ \\
    \hline
    ResBlock-3a & $\begin{bmatrix} 3 \times 3, 256\\ 3 \times 3, 256\\ 1 \times 1, 256 \end{bmatrix} \times 1$, stride=2 & $256 \times 20 \times T/4$ \\
    ResBlock-3b & $\begin{bmatrix} 3 \times 3, 256\\ 3 \times 3, 256 \end{bmatrix} \times 5$, stride=1 & $256 \times 20 \times T/4$ \\
    \hline
    ResBlock-4a & $\begin{bmatrix} 3 \times 3, 256\\ 3 \times 3, 256\\ 1 \times 1, 256 \end{bmatrix} \times 1$, stride=2 & $256 \times 10 \times T/8$ \\
    ResBlock-4b & $\begin{bmatrix} 3 \times 3, 256\\ 3 \times 3, 256 \end{bmatrix} \times 2$, stride=1 & $256 \times 10 \times T/8$ \\
    \hline
    Flatten $(C,F)$ & $-$ & $2560 \times T/8$ \\
    StatsPooling & $-$ & 5120 \\
    \hline
    Dense (Emb.) & $-$ & 128 \\
    Length-norm & $-$ & 128 \\
    Gauss quadratic-Softmax & $-$ & Num. spks. \\
    \bottomrule
  \end{tabular}
\end{table}

\subsection{Two Pass Approach}
We motivate the need for a two-pass enhancement to the LPG algorithm by noting that operating the LPG algorithm  does not work well if the segment length is small.  Long enough segments are needed to: 1) extract enough information from the speech segment to train an effective embedding, and 2) have enough stable SAD information to process the correct speech segments. However, longer segments translate to coarser speaker assignments in the diarization problem.  To enable fine temporal assignment of speakers to segments, we improve the LGP algorithm by adding a second pass.

The two pass modification works by running the LGP algorithm twice, with different segment lengths and overlaps.  The first pass is unchanged from our baseline approach, with a larger segment length (2 seconds) and no overlap. This configuration provides good performance overall. The second pass refines these diarization marks with a smaller segment length (1.25 sec) and high overlap (1 sec), to allow for temporal refinement of speaker assignments. We find that only 1 or 2 iterations are needed for this refinement process, as the clustering does not typically change very much.

Since the DNN is trained with particular target segment durations in mind, we find a small additional performance improvement by fine-tuning DNNs for each duration.
\subsection{Improved Duration Modeling}
\label{sec:neff}
\begin{figure}[t]
  \centering
  \includegraphics[width=\linewidth]{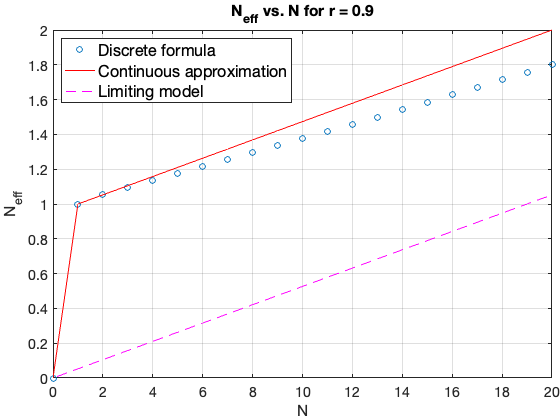}
  \caption{Comparison of functions for approximating effective number of samples.}
  \label{fig:Neff}
\end{figure}

We have addressed two areas of the correlation model for reducing effective counts and increasing uncertainty. First, the relation between the effective number of samples in a cluster and the actual count is a somewhat complicated discrete formula:
\begin{equation}
    \label{Neff_discr}
    N_{eff} = \frac{N}{1+2\sum_{j=1}^{N-1}{\frac{(N-j)}{N}r^j}}.
\end{equation}
In previous work, we have used the limiting approximation for large N of
\begin{equation}
    \label{Neff_lim}
    N_{eff} = \left(\frac{1-r}{1+r}\right) N.
\end{equation}
However, as shown in Fig.~\ref{fig:Neff}, this limiting function is not a close fit for small N with $r = 0.9$. Instead we introduce a more accurate continuous approximation:
\begin{equation}
    \label{Neff_cont}
    N_{eff} = \min\left(N,\frac{(1-r)N+2r}{1+r}\right).
\end{equation}
Using this continuous approximation in place of our previous interpolation of the discrete version is much simpler to implement with no loss in performance.

Secondly, we have found that diverse tasks in diarization can have wildly varying audio lengths. In particular, long audio files result in a large number of segments in each speaker cluster, which reduces uncertainty and does not match our DNN training conditions.
Therefore, we introduce a new parameter $N_0$ to represent the target number of segments in a file, and reduce the observed counts across longer files by the scale factor $N_0/N$.

\section{Experimental Results and Analysis}
\label{sec_exp}

We train the DNN  with LDC corpora Switchboard, Fisher, Mixer6, SRE2004-10, and VoxCeleb1~\cite{nagrani2017voxceleb}. With augmentations, this results in a set of 5,175,668 utterances from 13,129 speakers. We use a 90/10 split between training and development sets, yielding a training set of 11,816 speakers.  Segments vary between 1.5 and 2.5 seconds. Note that the DNN contains everything needed for diarization: both the speaker embedding (xvector) and internal PLDA parameters. No additional system training is needed.

\subsection{Callhome}
As our primary focus is telephone speech, we begin our evaluation of the two-pass LGP algorithm on the Callhome dataset.
Following standard practice, we report results for oracle speech activity marks with 250 ms forgiveness collar around speaker change points, and do not score segments with overlapping speakers.
The results are shown in Table~\ref{tab:callhome_results}, where the `System` column indicates the system being tested.  Here, we set $N_0=N$ since the number of segments in each audio file matches the DNN training conditions. The results show that the proposed two-pass LGP system exceeds the best current published performance.

\begin{table}[h!]
\centering
\caption{DER for various diarization algorithms for the Callhome dataset.}
\begin{tabular}{||c c||}
 \hline
 System & DER \\ [0.5ex]
 \hline\hline
  AHC & 8.10 \\
  AHC+VB~\cite{sell2018diarization} & 6.48 \\
  One-Pass LGP~\cite{mccree2019} & 6.62 \\
  VBx~\cite{landini2020bayesian} & 4.42 \\
  Two-Pass LGP & 3.92 \\ [1ex]
  \hline\hline
\end{tabular}
\label{tab:callhome_results}
\end{table}

In order to characterize the performance of each stage of the two-pass LGP algorithm, we also compute the DER after k-means initialization of the GMM clusters.  The DER decreases between the clustering initialization with k-means and the first pass by 81.9\%. The DER decreases between the first pass and the second pass by an additional 40.8\%.  The performance increase across each stage of processing indicates that while k-means is a good initializer of the GMM for clustering, by-itself, it does not achieve acceptable performance in assigning speakers to speech segments.  The first pass of assignment dramatically helps overall performance by refining initial speaker assignments provided by k-means.  As hypothesized, the second pass fine-tunes the speaker assignments successfully, and results in additional performance gain. With these results, we emphasize that the two-pass LGP algorithm used for the Callhome dataset is an integrated system and was not tuned on any development dataset.




\subsection{Wideband Datasets}
Next, we evaluate the proposed two-pass LGP algorithm on the wideband datasets: DIHARD2~\cite{ryant2019second}, AMI-Headset, and AMI-Beamformed~\cite{carletta2005ami}, using the same system as described above for Callhome. Here, we set $N_0=25$ to account for varying audio segment lengths as described in Section~\ref{sec:neff} for all evaluations conducted with the narrowband xvector system described in Section~\ref{sec_dnn_nb}. Because the DIHARD2 and AMI datasets are wideband, we also evaluate the two-pass LGP algorithm with an xvector system trained on wideband data. The wideband xvector system is trained with Voxceleb1~\cite{nagrani2017voxceleb} and Voxceleb2~\cite{chung2018voxceleb2}. With augmentations, this results in a set of 5,949,009 utterances from 6,825 speakers. We use a 90/10 split between training and development sets, yielding a training set of 6,143 speakers, with minibatches of size 128 with one segment per speaker.  The remainder of training is exactly the same as what was reported for the narrowband xvector system.  We set $N_0=30$ for the wideband xvector system, to account for varying audio segments.

We compare both the two-pass narrowband LGP algorithm (2P-NB-LGP) and the two-pass wideband LGP algorithm (2P-WB-LGP) to the VBx algorithm, using the task dependent tuning factors specified by the authors of that algorithm~\cite{landini2020bayesian}. For the DIHARD2 dataset, we evaluate using no collar and score overlapping speech segments.  For AMI, we follow the evaluation protocol specified by Section 4 in ~\cite{landini2020bayesian}, and show results for evaluating with no collar and scoring overlapping speech segments.

Table~\ref{tab:dihard_ami} aggregates the results, and shows that both two-pass LGP systems are competitive across all tasks.  Both the 2P-NB-LGP and 2P-WB-LGP exceed the best published DER performance for the evaluation split for both AMI-Beamformed (AMI-BF) and AMI-Mixed Headset (AMI-Mx) datasets.  The 2P-WB-LGP system shows competitive performance in the evaluation split of the DIHARD2 dataset. Using the $N_0$ robustness tuning factor, we revisit the performance of the 2P-NB-LGP algorithm on the Callhome dataset.  A configuration of $N_0=25$ reduces the DER performance on Callhome by 14.4\%.
We emphasize that while the NB and WB systems indicated by ``2P-NB-LGP'' and ``2P-WB-LGP'' in Table~\ref{tab:dihard_ami} are fundamentally different, each uses the same hyperparameters across the datasets tested.

\begin{table}[h!]
\centering
\caption{Diarization performance on DIHARD2 and AMI datasets}
\begin{tabular}{||c c||c c||}
 \hline
 Task  & System & Dev & Eval \\ [0.5ex]
 \hline\hline
 \multirow{6}{*}{DIHARD2} & Landini~\cite{landini2020but} & 17.90 & 18.21 \\
 & Lin~\cite{lin2020dihard} & 21.36 & 18.84 \\
 & Lin~\cite{lin2020self} & 18.76 & 18.44 \\
 & VBx & 18.19 & 18.55 \\
 & 2P-NB-LGP & 19.21 & 20.83 \\
 & 2P-WB-LGP & 17.8 & 18.76 \\ [1ex]
 \hline
 \multirow{3}{*}{AMI-BF} & VBx & 17.66 & 20.84 \\
 & 2P-NB-LGP & 18.56 & 19.95 \\
 & 2P-WB-LGP & 18.94 & 19.84 \\
 \hline
 \multirow{4}{*}{AMI-Mx} & VBx & 16.33 & 18.99 \\
 & 2P-NB-LGP & 16.21 & 17.57 \\
 & 2P-WB-LGP & 16.74 & 17.79 \\ [1ex]
 \hline
\end{tabular}
\label{tab:dihard_ami}
\end{table}

\section{Conclusion}

This paper has presented the first published number below 4\% DER for the Callhome benchmark task, without any task-dependent parameter estimation or tuning. This is achieved through significant improvements to the LGP diarization system, and particularly the two-pass approach to improve time resolution.  Additionally, since the LGP DNN is directly trained to optimize PLDA performance over segment embeddings, no further parameter training is needed to perform diarization. The single narrowband system also provides competitive performance across a range of tasks (Callhome, DIHARD2, and AMI). In addition, a single wideband version of the system can essentially match the tuned performance of the VBx approach over both DIHARD2 and AMI.

\eightpt
\bibliographystyle{IEEEtran}
\bibliography{speech,spkr,e2e_diarization}

\begin{thebibliography}{10}
\providecommand{\url}[1]{#1}
\csname url@samestyle\endcsname
\providecommand{\newblock}{\relax}
\providecommand{\bibinfo}[2]{#2}
\providecommand{\BIBentrySTDinterwordspacing}{\spaceskip=0pt\relax}
\providecommand{\BIBentryALTinterwordstretchfactor}{4}
\providecommand{\BIBentryALTinterwordspacing}{\spaceskip=\fontdimen2\font plus
\BIBentryALTinterwordstretchfactor\fontdimen3\font minus
  \fontdimen4\font\relax}
\providecommand{\BIBforeignlanguage}[2]{{%
\expandafter\ifx\csname l@#1\endcsname\relax
\typeout{** WARNING: IEEEtran.bst: No hyphenation pattern has been}%
\typeout{** loaded for the language `#1'. Using the pattern for}%
\typeout{** the default language instead.}%
\else
\language=\csname l@#1\endcsname
\fi
#2}}
\providecommand{\BIBdecl}{\relax}
\BIBdecl

\bibitem{sell2018diarization}
G.~Sell, D.~Snyder, A.~McCree, D.~Garcia-Romero, J.~Villalba, M.~Maciejewski,
  V.~Manohar, N.~Dehak, D.~Povey, S.~Watanabe \emph{et~al.}, ``Diarization is
  hard: Some experiences and lessons learned for the jhu team in the inaugural
  {DIHARD} challenge,'' in \emph{Proc. Interspeech}, 2018, pp. 2808--2812.

\bibitem{diez2018but}
M.~Diez, F.~Landini, L.~Burget, J.~Rohdin, A.~Silnova, K.~Zmol{\i}kov{\'a},
  O.~Novotn{\`y}, K.~Vesel{\`y}, O.~Glembek, O.~Plchot \emph{et~al.}, ``But
  system for {DIHARD} speech diarization challenge 2018,'' in \emph{Proc.
  Interspeech}, 2018, pp. 2798--2802.

\bibitem{sun2018speaker}
L.~Sun, J.~Du, C.~Jiang, X.~Zhang, S.~He, B.~Yin, and C.-H. Lee, ``Speaker
  diarization with enhancing speech for the first {DIHARD} challenge,''
  \emph{Proc. Interspeech 2018}, pp. 2793--2797, 2018.

\bibitem{zajic2018zcu}
Z.~Zaj{\i}c, M.~Kune{\v{s}}ov{\'a}, J.~Zelinka, and M.~Hr{\'u}z, ``Zcu-ntis
  speaker diarization system for the {DIHARD} 2018 challenge,'' in \emph{Proc.
  INTERSPEECH}, 2018, pp. 2788--2792.

\bibitem{vinals2018estimation}
I.~Vinals, P.~Gimeno, A.~Ortega, A.~Miguel, and E.~Lleida, ``Estimation of the
  number of speakers with variational bayesian plda in the {DIHARD} diarization
  challenge,'' in \emph{Proc. INTERSPEECH}, 2018, pp. 2803--2807.

\bibitem{patino2018eurecom}
J.~Patino, H.~Delgado, and N.~Evans, ``The eurecom submission to the first
  {DIHARD} challenge,'' in \emph{Proc. Interspeech}, vol. 2018, 2018, pp.
  2813--2817.

\bibitem{ryant2018first}
N.~Ryant, K.~Church, C.~Cieri, A.~Cristia, J.~Du, S.~Ganapathy, and
  M.~Liberman, ``First {DIHARD} challenge evaluation plan,'' 2018.

\bibitem{landini2020bayesian}
F.~Landini, J.~Profant, M.~Diez, and L.~Burget, ``Bayesian hmm clustering of
  x-vector sequences (vbx) in speaker diarization: theory, implementation and
  analysis on standard tasks,'' \emph{arXiv preprint arXiv:2012.14952}, 2020.

\bibitem{fujita2020end}
Y.~Fujita, S.~Watanabe, S.~Horiguchi, Y.~Xue, and K.~Nagamatsu, ``End-to-end
  neural diarization: Reformulating speaker diarization as simple multi-label
  classification,'' \emph{arXiv preprint arXiv:2003.02966}, 2020.

\bibitem{huang2020speaker}
Z.~Huang, S.~Watanabe, Y.~Fujita, P.~Garc{\'\i}a, Y.~Shao, D.~Povey, and
  S.~Khudanpur, ``Speaker diarization with region proposal network,'' in
  \emph{ICASSP 2020-2020 IEEE International Conference on Acoustics, Speech and
  Signal Processing (ICASSP)}.\hskip 1em plus 0.5em minus 0.4em\relax IEEE,
  2020, pp. 6514--6518.

\bibitem{medennikov2020target}
I.~Medennikov, M.~Korenevsky, T.~Prisyach, Y.~Khokhlov, M.~Korenevskaya,
  I.~Sorokin, T.~Timofeeva, A.~Mitrofanov, A.~Andrusenko, I.~Podluzhny
  \emph{et~al.}, ``Target-speaker voice activity detection: a novel approach
  for multi-speaker diarization in a dinner party scenario,'' \emph{arXiv
  preprint arXiv:2005.07272}, 2020.

\bibitem{kinoshita2020integrating}
K.~Kinoshita, M.~Delcroix, and N.~Tawara, ``Integrating end-to-end neural and
  clustering-based diarization: Getting the best of both worlds,'' \emph{arXiv
  preprint arXiv:2010.13366}, 2020.

\bibitem{horiguchi2020end}
S.~Horiguchi, Y.~Fujita, S.~Watanabe, Y.~Xue, and K.~Nagamatsu, ``End-to-end
  speaker diarization for an unknown number of speakers with encoder-decoder
  based attractors,'' \emph{arXiv preprint arXiv:2005.09921}, 2020.

\bibitem{mccree2019}
A.~McCree, G.~Sell, and D.~Garcia-Romero, ``Speaker diarization using
  leave-one-out gaussian plda clustering of dnn embeddings,'' in
  \emph{Interspeech}, 2019, pp. 381--385.

\bibitem{ioffe2006probabilistic}
S.~Ioffe, ``Probabilistic linear discriminant analysis,'' in \emph{European
  Conference on Computer Vision}.\hskip 1em plus 0.5em minus 0.4em\relax
  Springer, 2006, pp. 531--542.

\bibitem{Prince07}
S.~J.~D. Prince and J.~H. Elder, ``Probabilistic linear discriminant analysis
  for inferences about identity,'' in \emph{Proc. ICCV}, 2007, pp. 1--8.

\bibitem{villalba2015variational}
J.~Villalba, A.~Ortega, A.~Miguel, and E.~Lleida, ``Variational bayesian plda
  for speaker diarization in the mgb challenge,'' in \emph{2015 IEEE Workshop
  on Automatic Speech Recognition and Understanding (ASRU)}.\hskip 1em plus
  0.5em minus 0.4em\relax IEEE, 2015, pp. 667--674.

\bibitem{Duda01}
R.~O. Duda, P.~E. Hart, and D.~G. Stork, \emph{Pattern Classification}.\hskip
  1em plus 0.5em minus 0.4em\relax Wiley, 2001.

\bibitem{kenny2008bayesian}
P.~Kenny, ``Bayesian analysis of speaker diarization with eigenvoice priors,''
  \emph{CRIM, Montreal, Technical Report}, 2008.

\bibitem{diez2018speaker}
M.~Diez, L.~Burget, and P.~Matejka, ``Speaker diarization based on bayesian hmm
  with eigenvoice priors,'' in \emph{Proceedings of Odyssey}, 2018, pp.
  147--154.

\bibitem{Fukunaga90}
K.~Fukunaga, \emph{Introduction to Statistical Pattern Recognition}.\hskip 1em
  plus 0.5em minus 0.4em\relax Academic Press, 1990.

\bibitem{mccree2017extended}
A.~McCree, G.~Sell, and D.~Garcia-Romero, ``Extended variability modeling and
  unsupervised adaptation for plda speaker recognition.'' in
  \emph{Interspeech}, 2017, pp. 1552--1556.

\bibitem{snyder2019speaker}
D.~Snyder, D.~Garcia-Romero, G.~Sell, A.~McCree, D.~Povey, and S.~Khudanpur,
  ``Speaker recognition for multi-speaker conversations using x-vectors,'' in
  \emph{2019 IEEE International Conference on Acoustics, Speech and Signal
  Processing (ICASSP)}, 2019.

\bibitem{Romero11}
D.~Garcia-Romero and C.~Y. Espy-Wilson, ``Analysis of i-vector length
  normalization in speaker recognition systems,'' in \emph{Proc. Interspeech},
  2011, pp. 249--252.

\bibitem{garcia2020magneto}
D.~Garcia-Romero, G.~Sell, and A.~McCree, ``Magneto: X-vector magnitude
  estimation network plus offset for improved speaker recognition,'' in
  \emph{Proc. Odyssey 2020 The Speaker and Language Recognition Workshop},
  2020, pp. 1--8.

\bibitem{nagrani2017voxceleb}
A.~Nagrani, J.~S. Chung, and A.~Zisserman, ``Voxceleb: A large-scale speaker
  identification dataset,'' \emph{Proc. Interspeech 2017}, pp. 2616--2620,
  2017.

\bibitem{ryant2019second}
N.~Ryant, K.~Church, C.~Cieri, A.~Cristia, J.~Du, S.~Ganapathy, and
  M.~Liberman, ``The second dihard diarization challenge: Dataset, task, and
  baselines,'' \emph{arXiv preprint arXiv:1906.07839}, 2019.

\bibitem{carletta2005ami}
J.~Carletta, S.~Ashby, S.~Bourban, M.~Flynn, M.~Guillemot, T.~Hain, J.~Kadlec,
  V.~Karaiskos, W.~Kraaij, M.~Kronenthal \emph{et~al.}, ``The ami meeting
  corpus: A pre-announcement,'' in \emph{International workshop on machine
  learning for multimodal interaction}.\hskip 1em plus 0.5em minus 0.4em\relax
  Springer, 2005, pp. 28--39.

\bibitem{chung2018voxceleb2}
J.~S. Chung, A.~Nagrani, and A.~Zisserman, ``Voxceleb2: Deep speaker
  recognition,'' \emph{arXiv preprint arXiv:1806.05622}, 2018.

\bibitem{landini2020but}
F.~Landini, S.~Wang, M.~Diez, L.~Burget, P.~Mat{\v{e}}jka,
  K.~{\v{Z}}mol{\'\i}kov{\'a}, L.~Mo{\v{s}}ner, A.~Silnova, O.~Plchot,
  O.~Novotn{\`y} \emph{et~al.}, ``But system for the second dihard speech
  diarization challenge,'' in \emph{ICASSP 2020-2020 IEEE International
  Conference on Acoustics, Speech and Signal Processing (ICASSP)}.\hskip 1em
  plus 0.5em minus 0.4em\relax IEEE, 2020, pp. 6529--6533.

\bibitem{lin2020dihard}
Q.~Lin, W.~Cai, L.~Yang, J.~Wang, J.~Zhang, and M.~Li, ``Dihard ii is still
  hard: Experimental results and discussions from the dku-lenovo team,''
  \emph{arXiv preprint arXiv:2002.12761}, 2020.

\bibitem{lin2020self}
Q.~Lin, Y.~Hou, and M.~Li, ``Self-attentive similarity measurement strategies
  in speaker diarization,'' in \emph{Proc. Interspeech}, vol. 2020, 2020, pp.
  284--288.

\end{thebibliography}

\end{document}